\documentclass[preprint2]{aastex6}

\begin{document}

\title{Hypermassive Neutron Star Disk Outflows and Blue Kilonovae}

\author{Steven Fahlman and Rodrigo Fern\'andez}
\affil{Department of Physics, University of Alberta, Edmonton, AB T6G 2E1, Canada.}

\begin{abstract}
We study mass ejection from accretion disks around newly-formed
hypermassive neutron stars (HMNS). Standard kilonova model fits to GW170817
require at least a lanthanide-poor (`blue') and lanthanide-rich (`red') component. 
The existence of a blue component has been used as evidence for a HMNS remnant of 
finite lifetime, but average disk outflow velocities from existing long-term HMNS simulations 
fall short of the inferred value ($\sim 0.25c$) by a factor of $\sim 2$. Here we use time-dependent, 
axisymmetric hydrodynamic simulations of HMNS disks to explore the limits of the model 
and its ability to account for observations. For physically plausible parameter choices 
compatible with GW170817, we find that hydrodynamic models that use shear viscosity to 
transport angular momentum cannot eject matter with mass-averaged velocities larger than 
$\sim 0.15c$. While outflow velocities in our simulations can exceed the asymptotic value
for a steady-state neutrino-driven wind,
the increase in the average velocity due to viscosity is not sufficient. 
Therefore, viscous HMNS disk winds cannot reproduce by themselves the ejecta properties inferred 
from multi-component fits to kilonova light curves from GW170817. 
Three possible resolutions remain feasible within standard merger ejecta channels: 
more sophisticated radiative transfer models that allow for photon reprocessing between ejecta
components, inclusion of magnetic stresses, or enhancement of the dynamical ejecta.
We provide fits to our disk outflow models once they reach homologous expansion.
\end{abstract}


\section{Introduction} 
\label{sec:intro}

The optical and infrared emission accompanying the neutron star (NS)
merger GW170817\footnote{Also known as GRB170817A, SSS17a, AT 2017gfo, and DLT17ck.} 
\citep{ligogw170817multi-messenger} 
is broadly consistent with the predictions of
the \emph{kilonova/macronova} model: a thermal transient powered by
the radioactive decay of $r$-process elements
on sub-relativistic ejecta from the merger \citep{LP1998,Metzger2010,tanaka2016}. 
This agreement positions NS mergers as an important astrophysical
site for the $r$-process (e.g., \citealt{cote2018}).

The transient initially peaked in the optical/UV, 
transitioning to the near-infrared within a few days (e.g., 
\citealt{Cowperthwaite2017,drout2017,tanaka2017}).
This behavior is consistent with some of the ejecta having a large
opacity due to the presence of lanthanides and/or 
actinides \citep{Kasen2013,Tanaka2013,fontes2015}.

The most common approach to infer the ejecta properties 
is to fit at least two kilonova components
that evolve independently (e.g., \citealt{kasen2017}). This approach yields 
a lanthanide-rich (`red') component expanding at $\sim 0.15c$ with mass in the 
range $0.02-0.05M_\odot$, and a faster lanthanide-poor (`blue') kilonova moving 
at $\sim 0.2-0.3c$ and with mass $\lesssim 0.02M_\odot$ 
(e.g., \citealt{Villar2017} and references therein).

Theoretically, NS-NS mergers can generate multiple
ejecta components. For the inferred parameters of GW170817, numerical
relativity simulations predict dynamical ejecta masses $\lesssim 0.01M_\odot$,
with velocities of $0.2-0.3c$, and a mostly lanthanide-rich composition, while
the remnant accretion disk is expected to have masses in the range $0.05-0.3M_\odot$
depending on the equation of state (e.g., \citealt{Shibata2017}). The disk can eject a substantial
fraction of its mass over timescales much longer than the dynamical time \citep{FM13,Just2015}.
Therefore, the dominant ejection channel by mass for NS-NS mergers compatible with 
GW170817 is expected to be the disk outflow.

If a hypermassive NS (HMNS) survives for longer than the dynamical time, the disk 
can eject a significant amount of lanthanide-poor material due to the enhanced neutrino reprocessing of
the ejecta (\citealt{MF14}, hereafter MF14; \citealt{perego2014,Lippuner2017,fujibayashi2018}) and because the disk
mass itself is larger than the case in which the black hole (BH) forms promptly (e.g., \citealt{Hotokezaka2013}). The presence
of blue optical emission in the kilonova has been used as evidence for a
HMNS remant of finite lifetime in GW170817 (e.g., \citealt{Bauswein2017,Margalit2017}).

Existing simulations of the long-term evolution of disks around HMNS remnants
are few and all hydrodynamic, either using parameters
not directly applicable to GW170817 (MF14; \citealt{Lippuner2017}), not including viscous angular 
momentum transport explicitly \citep{Dessart2009,perego2014}, or never collapsing into black 
holes \citep{fujibayashi2018}. In addition, the disk outflow in all these simulations achieves
mass-averaged velocities of at most $\sim 0.1c$, which is significantly
slower than the inferred blue kilonova component (e.g., \citealt{Metzger2018}).

Here we revisit mass ejection from HMNS disks using axisymmetric
hydrodynamic simulations that approximate the dominant physical effects.
These include neutrino irradiation by the HMNS, freezout of weak interactions
in the disk on the viscous timescale, and energy deposition by nuclear recombination
and (turbulent) angular momentum transport. Our aim is to compare the 
properties of the resulting ejecta with that inferred from observations, thereby exploring
the limits of the HMNS disk outflow model given the physics included. 
In doing so, we parameterize our ignorance about some
effects (e.g., lifetime of the HMNS) and our approximate modeling of
others (e.g., angular momentum transport), using plausible choices for input parameters
that are also compatible with GW170817. 

\section{Methods}
\label{sec:methods}

Disk outflow simulations use the same approach as in MF14, with updates
reported in \citet{Lippuner2017}. Below is a brief summary of the
computational setup.

\subsection{Numerical Hydrodynamics}

Simulations are carried out using {\tt FLASH3} \citep{fryxell00,Dubey2009} with
suitable modifications (\citealt{FM13}; MF14). The code solves
the equations of hydrodynamics and lepton number conservation
in axisymmetric (2D) spherical polar coodinates ($r,\theta$) with azimuthal rotation.
Gravity, azimuthal shear viscosity, and neutrino emission/absorption
are included as source terms. We use the equation of state of
\citet{TimmesSwesty2000} with abundances of neutrons, protons,
and alpha particles in nuclear statistical equilibrium, and accounting for
the nuclear recombination energy of alpha particles.

Gravity is modeled with the pseudo-Newtonian potential of \citet{Artemova1996},
azimuthal shear viscosity follows an $\alpha$-prescription \citep{shakura1973}, and neutrino effects
are modeled with a leakage scheme for emission and annular light bulb
for absorption (\citealt{FM13}; MF14). We only include charged-current weak interactions
on nucleons. See \citet{richers2015} for a comparison
of this scheme with Monte Carlo neutrino transport.

The computational domain is discretized radially
using logarithmic spacing with $128$ cells per decade in radius, and 
using 112 cells equispaced in $\cos\theta$ covering the range $[0,\pi]$. 

The HMNS is modeled as a reflecting inner radial boundary at $r=R_{\rm NS}$,
from which prescribed neutrino and antineutrino luminosities are emitted. These
luminosities are constant for the first $10$\,ms, subsequently decaying as
$t^{-1/2}$ (MF14). 
When the HMNS collapses into a BH, the radial boundary becomes absorbing,
and the HMNS luminosities are set to zero. The boundary is also
moved inward to a position halfway between the innermost stable circular orbit (ISCO) and horizon radii of the
newly-formed BH. The computational domain extends out to $r=2\times 10^9$\,cm.
The outer radial boundary condition is absorbing, and the boundary conditions
in $\theta$ are reflecting.

\begin{table*}

\caption{Simulation parameters and results. Columns from left to right show model name,
central object mass, HMNS radius, initial torus mass,
radius of initial torus density maximum, initial HMNS neutrino luminosity ($L_{\nu_e}=L_{\bar\nu_e}$), 
initial torus electron fraction, HMNS lifetime,
viscosity parameter, initial torus entropy,
ejected mass with positive energy in lanthanide-rich
($Y_e < 0.25$, subscript R for red) and lanthanide-poor ($Y_e > 0.25$, subscript B for blue)
material, and mass-averaged velocity of ejected red and blue material.
}
\centering
\label{tab:simoutcomes}
\begin{tabular}{l|ccccccccc|cccc}
 Model & $M_{\rm NS}$ & $R_{\rm NS}$ & $M_{\rm t}$ & $R_{\rm t}$  & $L_{\nu e}$ & $Y_e$ & $\tau_{\rm NS}$ & 
   $\alpha$ & s & $ \bar{\rm M}_{\rm R} $ & $\bar{\rm M}_{\rm B}$ & $\bar{\rm v}_{\rm R}$ & $\bar{\rm v}_{\rm B}$  \\ 

  & ($M_{\odot}$) & (km) & ($M_{\odot}$) & (km) & (ergs) &  & (ms) &  & ($k_{\rm B}$/baryon) & ($M_{\odot}$) & 
       ($M_{\odot}$) & (c) & (c) \\ \hline \hline

 base        & 2.65 & 20 & 0.10 & 50 & $2 \cdot 10^{52}$ & 0.10 & 10  & 0.05 & 8 &0.010 & 0.023 & 0.091 & 0.038 \\ 

 $\alpha$10  &      &    &      &    &                  &      &     & 0.10 &   & 0.008 & 0.035 & 0.135 & 0.070 \\ 
 $\alpha$03  &      &    &      &    &                  &      &     & 0.03 &   & 0.007 & 0.019 & 0.066 & 0.032 \\  
 t01         &      &    &      &    &                  &      & 1   &      &   & 0.013 & 0.008 & 0.037 & 0.039 \\ 
 
 t30         &      &    &      &    &                  &      & 30  &      &   & 0.002 & 0.058 & 0.159 & 0.093 \\  

 M2.7        & 2.70 &    &      &    &                  &      & 10  & 0.05 &   & 0.009 & 0.023 & 0.097 & 0.042 \\  
 M2.6        & 2.60 &    &      &    &                  &      &     &      &   & 0.011 & 0.018 & 0.080 & 0.041 \\ 
 
 mt03        & 2.65 &    & 0.30 &    &                  &      &     &      &   & 0.049 & 0.031 & 0.049 & 0.039 \\  
 mt02        &      &    & 0.20 &    &                  &      &     &      &   & 0.029 & 0.033 & 0.065 & 0.030 \\
 
 rt60        &      &    & 0.10 & 60 &                  &      &     &      &   & 0.014 & 0.013 & 0.057 & 0.039 \\ 
 rs30        &      & 30 &      & 50 &                  &      &     &      &   & 0.016 & 0.009 & 0.042 & 0.041 \\ 
  
 L53         &      & 20 &      &    & $2\cdot 10^{53}$ &      &     &      &   & 0.001 & 0.041 & 0.187 & 0.099 \\ 
 L51         &      &    &      &    & $2\cdot 10^{51}$ &      &     &      &   & 0.013 & 0.017 & 0.077 & 0.039 \\ 
 
 s10         &      &    &      &    & $2\cdot 10^{52}$ &      &     &      & 10 & 0.020 & 0.014 & 0.055 & 0.033 \\ 
 
 ye25        &      &    &      &    &                  & 0.25 &     &      & 8  & 0.000 & 0.033 & 0.000 & 0.058 \\ 
 best        & 2.55 & 20 & 0.20 & 60 & $2\cdot 10^{52}$ & 0.10 & 10  & 0.05 & 8  & 0.040 & 0.022 & 0.043 & 0.037   
\end{tabular}
\end{table*}

The initial condition for the disk is an equilibrium torus with constant
angular momentum, entropy, and electron fraction. The space outside
this torus is filled with an inert low-density ambient medium with
density in the range $10-100$~g\,cm$^{-3}$ inside $r=2\times 10^7$\,cm, and
decreasing as $r^{-2}$ outside this radius. When collapsing the HMNS into a BH, the 
cells added to the computational domain are filled with material having the same properties 
as the surrounding medium, which is immediately accreted. For numerical reasons, 
we set a floor of density at $\sim 90\%$ of the initial ambient value.

\subsection{Model Parameters}

The total mass of GW170817 measured from gravitational waves is
$2.73^{+0.04}_{-0.01}\,M_\odot$ to 90\% confidence \citep{LVSC2017a}. The dynamical ejecta
mass expected from numerical relativity simulations is $\lesssim 0.01M_\odot$, and disk
masses are expected to lie in the range $0.05-0.3M_\odot$ depending on the
equation of state used (e.g., \citealt{Shibata2017}). We therefore adopt a baseline
model (`base') with HMNS mass $M_{\rm NS} = 2.65M_\odot$ and disk mass $M_{\rm t} = 0.1M_\odot$.

The radius of the baseline HMNS is taken to be $R_{\rm NS}=20$\,km, following results
of numerical relativity simulations (e.g., \citealt{hanauske2017,Shibata2017b}).
The lifetime of the baseline HMNS is taken to be $\tau_{\rm NS} = 10$\,ms as a first guess ($\sim$ disk thermal time),
with the HMNS luminosities having an initial magnitude $2\times 10^{52}$\,erg\,s$^{-1}$
(e.g., \citealt{Dessart2009}). The HMNS has a surface rotation period $1.5$\,ms and we adopt zero spin
in the pseudo-Newtonian potential. The HMNS collapses into 
a BH of the same mass and dimensionless spin $0.8$, as typically obtained in numerical relativity simulations 
(e.g., \citealt{Shibata2017}). The inner radial boundary then moves from $20$\,km to $8.7$\,km in the baseline model.
The magnitude of the $\alpha$-viscosity
is chosen to be $\alpha=0.05$, following the GRMHD results of \citet{Fernandez2018}.
The initial electron fraction and entropy of the baseline disk are $Y_e = 0.1$ and $s=8$\,k$_{\rm B}$
per baryon, respectively. All model parameters are summarized in Table~\ref{tab:simoutcomes}.

We evolve additional models that vary one parameter at a time relative to the
baseline simulation, as shown in Table~\ref{tab:simoutcomes}. 
We focus on those parameters that are known to have the
most impact in the properties of the outflow: lifetime of the HMNS, magnitude
of the $\alpha$-viscosity, magnitude of the HMNS luminosity, 
mass of the torus and total remnant mass, and radius of the HMNS. 
Other parameters have a smaller impact on the disk evolution (\citealt{FM13}; MF14). 


\begin{figure*}
\includegraphics*[width=\textwidth]{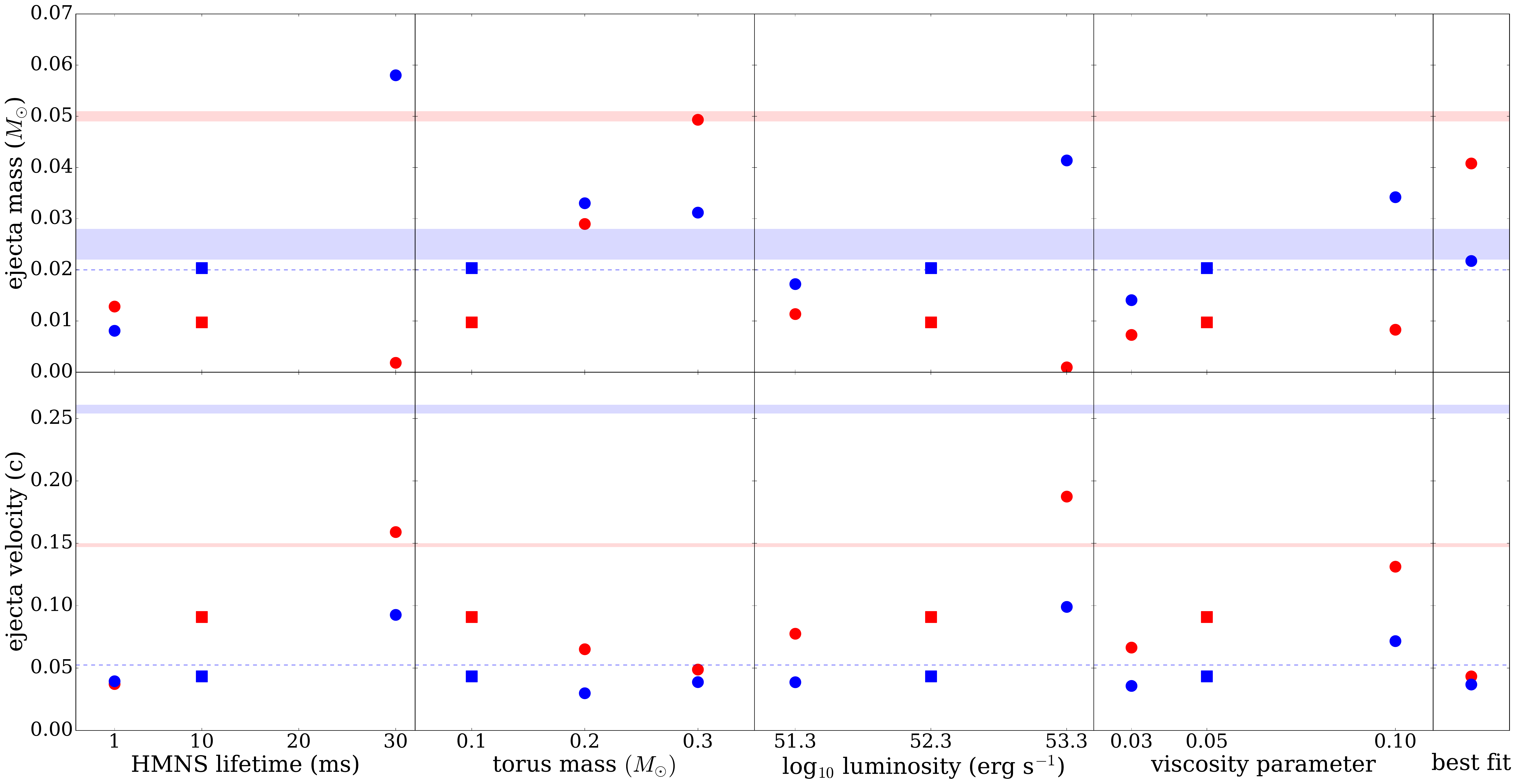}
\caption{Ejected mass (top) and mass-averaged velocity (bottom) of the unbound disk outflow as a function of
selected parameters: HMNS lifetime, initial torus mass, initial HMNS neutrino/antineutrino luminosity, 
and $\alpha$-viscosity parameter. The rightmost panel shows our `best fit' model. Red and blue symbols
denote lanthanide-rich ($Y_e < 0.25$) and poor material, respectively, with
squares denoting our `base' model (Table~\ref{tab:simoutcomes}).
Data points have a fiducial uncertainty of $10\%$ due to resolution.
The horizontal red and blue bands correspond to the two-component fit of GW170817 by \citet{Villar2017}, while
the horizontal dotted blue line corresponds to the (average) disk outflow values used
by \citet{Kawaguchi2018}. 
}
\label{fig:model_fits}
\end{figure*}

\section{Results}    
\label{sec:results}

\subsection{Overview of Disk Evolution}
\label{sec:evo}

The qualitative evolution of the torus is independent of 
parameter choices, for details see MF14 and \citet{Lippuner2017}.
While the HMNS is present, accretion of the disk forms a
high-density ($\sim 10^{12} \; \rm{g \; cm}^{-3}$) boundary layer around the
HMNS. Due to intense neutrino and viscous heating, material is ejected from the 
boundary layer and from the edges of the disk on the local thermal time ($\sim 10$\,ms). 
Material escaping within $\sim 20$\,deg of the polar axis has $Y_e\sim 0.5$ due
to strong irradiation, while on the equator the outflow has a $Y_e$ closer to
the initial disk value. The bulk of the disk remains neutron-rich ($Y_e\sim 0.2$) 
due the higher densities and shadowing of neutrino irradiation.    

Upon collapse of the HMNS into a BH,
the boundary layer accretes within $\sim 0.1$ ms, and a rarefaction wave is launched
outward. The torus readjusts on the equatorial
plane, evacuating the polar funnel. After a viscous time ($\sim 100-300$\,ms),
weak interactions freeze out and mass is ejected due to heating by
viscosity and nuclear recombination. By this time the electron fraction of the outflow is
higher than the initial disk value due to the lower degeneracy ($Y_e\sim 0.2-0.3$).

\subsection{Parameter Sensitivity}

Table~\ref{tab:simoutcomes} shows the mass and 
mass-averaged radial velocity of unbound disk ejecta for all models, as measured at a radius $r=10^9$\,cm. 
We use $Y_e=0.25$ to divide the ejecta into lanthanide-poor (`blue') and
lanthanide-rich (`red') material (e.g., \citealt{Lippuner2015}). 
Figure~\ref{fig:model_fits} illustrates the most sensitive parameter
dependencies. While here we use the 
two-component fit of \citet{Villar2017} as a reference observational result,
our general conclusions are independent of the specific (multi-component) kilonova fit used.

Our baseline model ejects an amount of mass with $Y_e > 0.25$ that approaches the
observationally-inferred value, but there is insufficient lanthanide-rich mass ejected by a factor of $5$. 
Also, the average velocity of the blue component is lower than that of the red ejecta, 
with the latter being $0.09c$ only.

The larger amount of blue relative to red ejecta for the default
HMNS lifetime ($\sim 10$\,ms) differs from that obtained by MF14, because the
latter used a non-spinning BH after HMNS collapse. 
The red ejecta is produced in the initial thermal expansion of the disk on
the side of the torus opposite to the HMNS,
before weak interactions have time to significantly reprocess the 
disk composition, and therefore depends entirely on the initial condition chosen 
in the baseline model ($Y_e=0.1$). Model ye25 imposes $Y_e = 0.25$ initially
in the disk, resulting in negligible red ejecta. 

\begin{figure*}
\centering
\includegraphics*[width=0.8\textwidth]{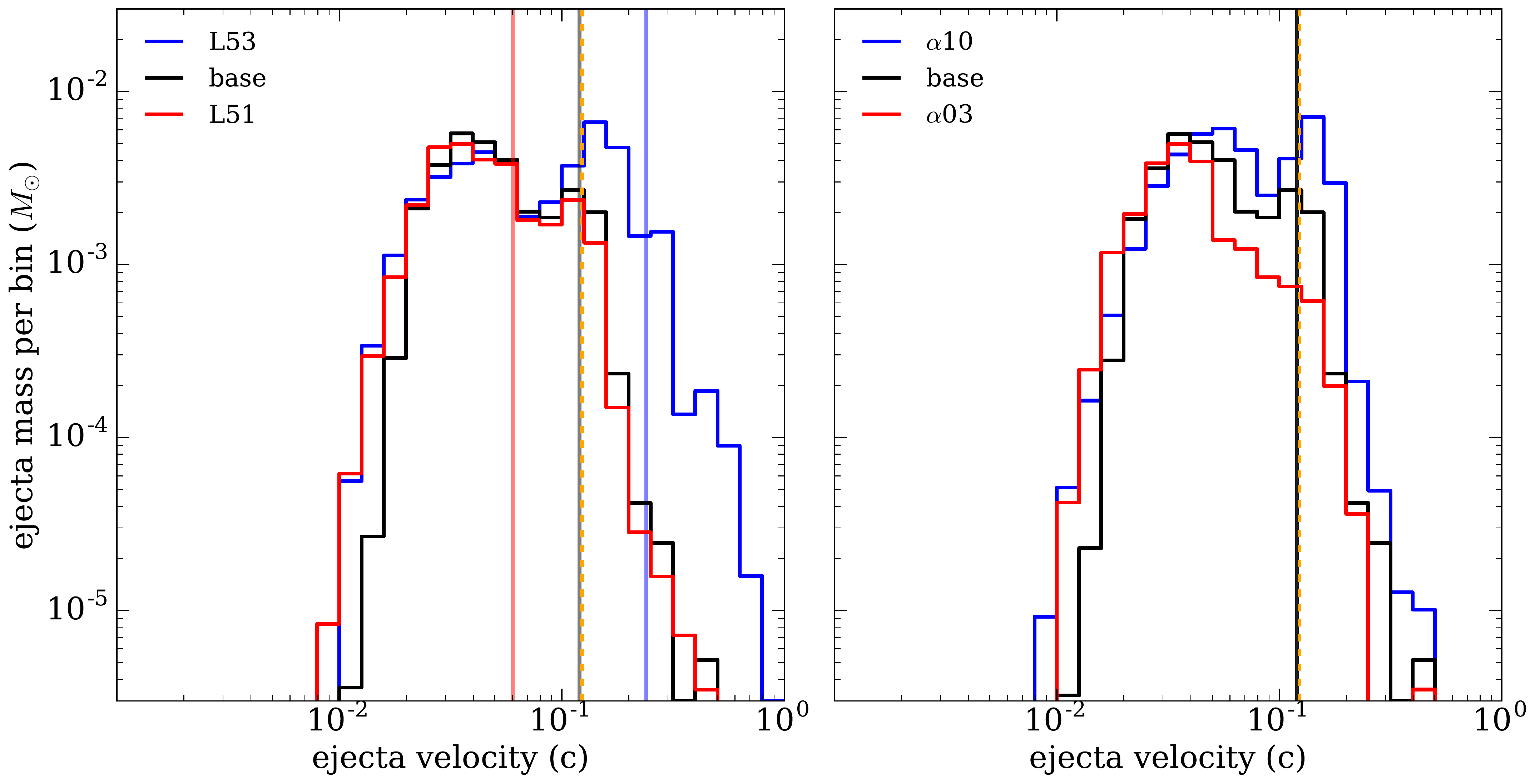}
\caption{Mass histogram of unbound material ejected at $r=10^9$\,cm, as a function of velocity, for
models that vary the magnitude of the initial HMNS luminosity (left) and the viscosity parameter (right), as labeled.
The vertical solid lines show the asymptotic velocity for a pure neutrino driven wind 
(equation~\ref{eq:v_neutrino_thompson}), and the vertical dashed line shows the maximum 
velocity achievable from alpha particle recombination energy alone (equation~\ref{eq:v_alpha}).
}
\label{fig:histograms}
\end{figure*}

Increasing the HMNS lifetime, viscosity parameter, or initial HMNS
luminosity results in the same trend: higher blue mass ejected, constant or
decreasing red mass, and moderate increase in the outflow velocities. In all
cases, the average velocity of the blue ejecta does not exceed $0.1c$, and
the red ejecta exceeds $0.15c$ only when its mass is $\ll 0.01M_\odot$. The physics
behind this trend is different in each case: longer HMNS lifetime results
in longer neutrino irradiation and absence of mass/energy loss to the BH,
higher viscosity parameter increases viscous heating (thereby increasing
the entropy and thus equilibrum $Y_e$) and accelerates the
disk evolution, while a higher HMNS luminosity increases the strength
of neutrino heating and accelerates the rate of change of $Y_e$.

Increasing the torus mass increases the lanthanide-rich mass, in part due
to a larger thermal outflow that contains the most neutron-rich material,
but also because the late-time viscous outflow becomes more neutron-rich.
The blue mass peaks at $M_t = 0.2M_\odot$ and then decreases for higher tori masses.
The average velocities of both components remain below $0.05c$.

Changes in the initial torus properties other than mass or composition produce
minor quantitative changes, as illustrated by models rt60 and s10. Similarly, changes in the
mass of the central object yield the same qualitative result.
Increasing the HMNS radius 
increases the surface area of the star 
and decreases the density in the boundary layer, resulting in
stronger torus irradiation and thus a higher electron fraction in the outflow. 
However, the total ejected mass is not significantly affected. We caution that 
these effects may be unique to our treatment of the HMNS as a hard boundary.

Finally, our best fit model involves increasing the torus mass and formation
radius. The combination of these effects creates outflows with red and blue
masses close to observational fits (allowing for an additional $0.01M_\odot$
supplement of red dynamical ejecta), but with lower average velocities for the blue
component than required. 

\begin{figure*}
\centering
\includegraphics[width=\textwidth]{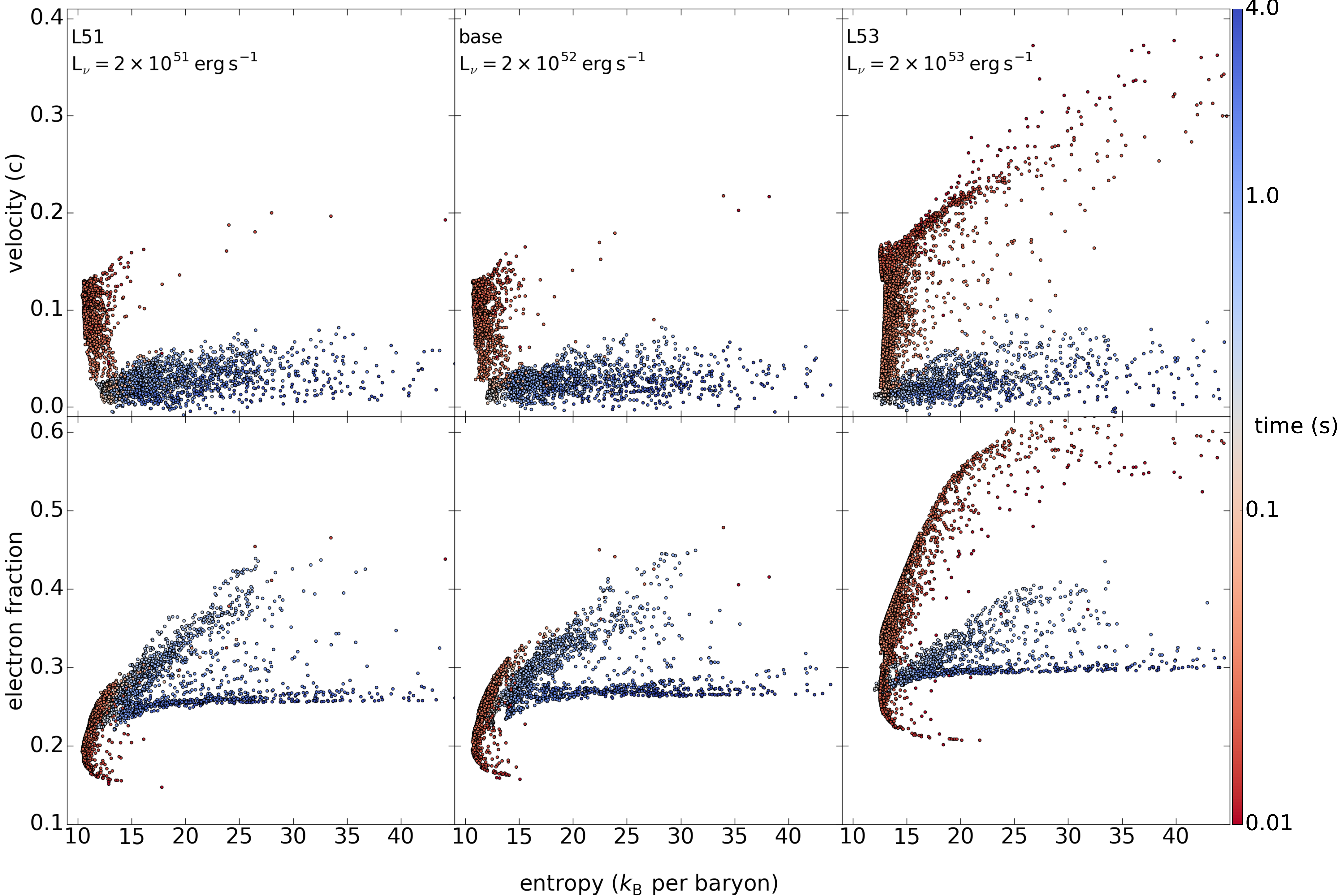}
\caption{Properties of tracer particles ejected with positive energy, for models that vary
the HMNS luminosity, as labeled, illustrating the different components of the disk outflow.
The color of each particle corresponds to the time at which
the temperature is $T=5\times 10^9$\,K for the last time before ejection. The
velocity, electron fraction, and entropy shown are the values attained at this time.
}
\label{fig:nu_wind}
\end{figure*}

\subsection{Physical constraints on the outflow velocity}

The initial thermal outflow is launched by a combination of viscous
and neutrino heating. Viscous angular momentum transport enhances
the outflow relative to a pure neutrino driven wind, by transporting
material to shallower regions of the potential well, in addition to
enhancing energy deposition \citep{Lippuner2017}. 

Figure~\ref{fig:histograms} illustrates the magnitude of this effect for
models that vary the neutrino luminosity and viscosity parameter.
The velocity distribution of the outflow is broad, and always
exceeds the asymptotic velocity obtained in
steady-state neutrino-driven wind models
\citep{Thompson2001,Metzger2018}
\begin{equation}
\label{eq:v_neutrino_thompson}
v_\nu \approx 0.12c \bigg{(} \frac{L_{\nu}}{2\times10^{52}}\bigg{)}^{0.3}.
\end{equation}  
The low-velocity portion of the distribution
is ubiquitous to all models, arising from the late-time 
viscous/recombination-driven outflow which is launched once the disk 
has spread to larger radii. This component has an upper limit
close to the maximum velocity that can be gained from nuclear recombination of alpha particles
\begin{equation}
\label{eq:v_alpha}
v_{\rm rec} = \sqrt{2 B_{\alpha}/m_{\alpha}} \approx 0.12c,
\end{equation}
where $B_\alpha$ and $m_\alpha$ are nuclear binding energy and mass
of an alpha particle (see, e.g., \citealt{Fernandez2018}).

The different components of the outflow can be separated with
tracer particles \citep{Lippuner2017}, as shown in Figure~\ref{fig:nu_wind}
for the models that vary the neutrino luminosity.
The prompt ($t<0.1$\,s) neutrino-driven wind appears as a tight correlation between
the entropy and electron fraction of the particles. The importance
of this component increases significantly with increasing neutrino
luminosity, with the correlation extending to higher velocities
and electron fractions. An intermediate component ($0.1<t<1$\,s)
also shows a correlation between entropy and electron fraction extending up to $Y_e = 0.4$, 
but with a larger scatter than the prompt outflow and a lower velocity ($<0.1c$). 
The late-time viscous/recombination-powered wind in the advective phase
($t>1$\,s) has nearly constant average velocity ($\lesssim 0.05c$) 
and electron fraction ($\lesssim 0.3$), 
but with a wide range of entropies.

Out of these components, only the prompt viscously-enhanced neutrino-driven 
wind is able to significantly exceed $0.1c$. However, in our most extreme case (model L53), 
the ejected mass with speeds above $0.2c$ and 
$Y_e > 0.25$ is less than $3 \times 10^{-3}M_\odot$. 

We conclude that a combination of neutrino heating and viscous angular momentum
transport in hydrodynamics is not able to account for the observed
components of the GW170817 when considering the HMNS disk outflow alone. This
conclusion is not altered by our omission of full general relativistic
effects, since the dynamics close to the BH horizon is not a key element for the
generation of outflows while the HMNS is present, and our results are consistent with those 
of \citet{fujibayashi2018}, who include all relativistic effects.

\subsection{Homologous disk ejecta}

\begin{figure*}
\begin{minipage}{0.9\columnwidth}

\includegraphics*[width=\columnwidth]{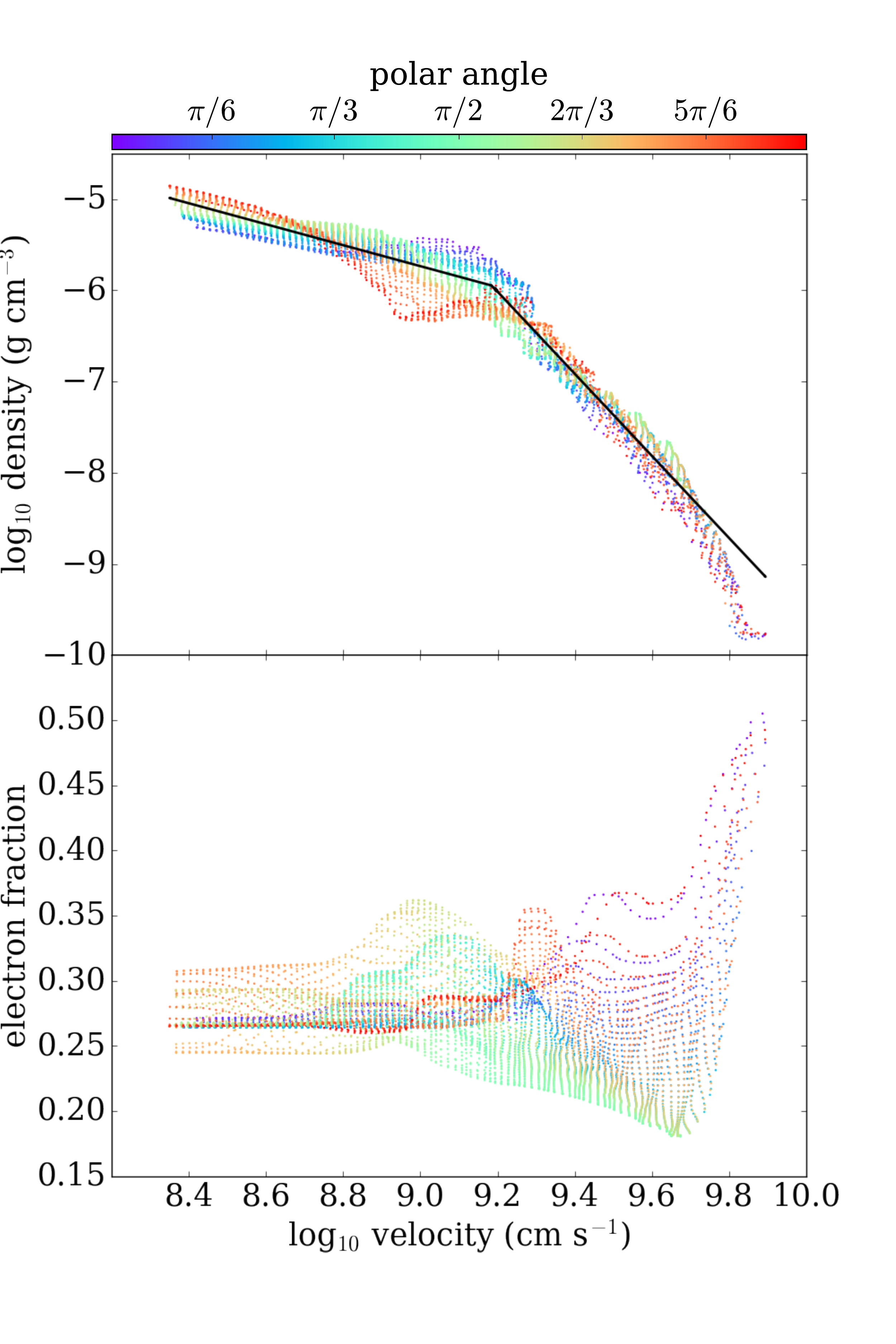}
\end{minipage}
\begin{minipage}{\columnwidth}
\centering
\begin{tabular}{l|cccccc|c}
Model & $\rho_0/10^{-5}$ & $v_0/c$ & $v_1/c$ & $v_2/c$ & $\eta_0$ & $\eta_1$ & $\bar{Y}_e$ \\ 
      & (g\,cm$^{-3}$) & ($10^{-3}$) & ($10^{-2}$) & ($10^{-1}$) & & \\ \hline \hline
base & 1.024 & 7.479 &  5.101 & 2.604  & 1.150  & 4.503 & 0.264 \\  
a10  & 0.745 & 7.771 &  6.119 & 2.872  & 1.125  & 3.678 & 0.287 \\ 
a03  & 4.485 & 6.955 &  3.827 & 2.371  & 1.788  & 4.858 & 0.260 \\ 
t01  & 2.414 & 7.172 &  5.347 & 1.443  & 1.665  & 7.671 & 0.238 \\ 
t30  & 0.585 & 7.739 &  5.337 & 3.146  & 0.875  & 2.946 & 0.354 \\ 
M2.7 & 2.195 & 7.138 &  4.619 & 2.696  & 1.479  & 4.154 & 0.269 \\ 
M2.6 & 2.430 & 7.109 &  6.632 & 2.459  & 1.741  & 5.276 & 0.257 \\
L51  & 1.893 & 7.488 &  5.234 & 2.314  & 1.525  & 4.708 & 0.254 \\ 
L53  & 1.667 & 7.535 &  4.728 & 4.349  & 1.512  & 3.252 & 0.354 \\ 
mt03 & 6.673 & 6.828 &  4.586 & 2.263  & 1.404  & 5.266 & 0.233 \\ 
mt02 & 3.714 & 6.822 &  3.461 & 2.501  & 0.996  & 4.376 & 0.242 \\ 
rt60 & 3.551 & 6.868 &  5.747 & 2.251  & 1.818  & 5.589 & 0.242 \\
rs30 & 2.543 & 7.214 &  4.126 & 2.019  & 1.445  & 5.197 & 0.233 \\ 
s10  & 1.601 & 7.414 &  4.352 & 2.366  & 1.125  & 4.936 & 0.235 \\ 
ye25 & 1.299 & 7.463 &  4.091 & 2.482  & 1.095  & 4.010 & 0.311 \\ 
best & 6.284 & 6.623 &  4.597 & 2.415  & 1.354  & 5.673 & 0.233  
\end{tabular}

\end{minipage}
\caption{\emph{Left:} Ejecta in the homologous phase for the base model ($t=1000$s), shown as
density (top) and electron fraction (bottom) as a function of radial velocity
in each computational cell, colored by polar angle. The solid line shows
a broken power-law fit to the density profile.
\emph{Right:} Parameters of the broken power-law fit to the density
in homology (equation~\ref{eq:homology_fit}). For reference, 
we also provide the mass-averaged electron fraction of the outflow.}
\label{tab:homology}
\end{figure*}

For reference, we provide fits to our disk ejecta once it has reached homologous
expansion, as needed for radiative transfer models. We compute the evolution into this 
phase ($\sim 1000$\,s after merger)
following the same method as in \citet{Kasen2015}. Figure~\ref{tab:homology} 
shows the density and electron fraction profiles for the baseline model in this phase.
For the ejecta density, we obtain acceptable fits with a broken power-law over a finite velocity range:
\begin{equation}
\rho/\rho_0 = \Bigg\{
\begin{array}{ll}
(v/v_0)^{-\eta_0} & v_0 < v < v_1\\
(v_1/v_0)^{-\eta_0}\,(v/v_1)^{-\eta_1} & v_1 < v < v_2,
\end{array}
\label{eq:homology_fit}
\end{equation}
where $\rho$ and $v$ are the ejecta density and radial velocity, respectively.
The velocity range $[v_0,v_2]$ is fixed by requiring that $90\%$ of the 
energy is kinetic, and it is beyond the turbulent region ($r > 1.26 \times 10^6$ km). The remaining variables $(\rho_0,v_1,\eta_0,\eta_1)$ are fit parameters. 

The electron fraction has a more complicated behavior,
hence we do not attempt to fit it. Parameters for equation~(\ref{eq:homology_fit})
and average electron fraction are given in the right panel of Figure~\ref{tab:homology}.

\section{Summary and Discussion} 
\label{sec:conc}

We have studied the long-term outflows from disks around HMNS remnants that 
collapse into BHs, using axisymmetric hydrodynamic simulations that include
the dominant physical effects save for magnetic stresses. We find that
for plausible parameters compatible with GW170817, hydrodynamic disk
outflow models that employ shear viscosity to transport angular momentum
cannot achieve mass-averaged velocities compatible with the blue
kilonova as inferred from multi-component kilonova fits. While the ejected mass
can in principle be brought closer to the inferred values by a suitable parameter choice,
the same cannot be achieved for the velocities of both components.

\citet{Kawaguchi2018} find that radiative transfer simulations that include
reprocessing of photons from the disk outflow by the dynamical ejecta do not
require a disk wind expanding faster than $0.1c$ to explain the GW170817 kilonova. 
Here the dynamical ejecta provides a velocity boost to these blue photons,
and eliminates the need for high ejecta masses, bringing it into agreement
with current predictions from numerical relativity simulations. Our disk outflow
models are fully compatible with their results (c.f. Figure~\ref{fig:model_fits}).
Establishing whether this is the correct resolution to the wind velocity problem 
requires further work.

Alternatively, state-of-the-art numerical relativity simulations predict too little dynamical
ejecta to reconcile the large masses moving at $0.25c$. Enhancements in this
prompt ejecta can be obtained for example by viscous effects, either by ejecting
material directly from the HMNS at early times \citep{Shibata2017c}, or by thermally 
boosting the dynamical ejecta \citep{radice2018}. The robustness of these
effects remains to be further explored.

The only remaining way to significantly boost the disk velocities are magnetic stresses.
Initial three-dimensional GRMHD models of BH remnant disks show that this
can easily be achieved \citep{Siegel2018,Fernandez2018}. The conjecture is further supported by
early-phase simulations of magnetized, differentially rotating HMNS remnants
(e.g., \citealt{Kiuchi2012,Siegel2014}). Including the effects of magnetic
fields is the most straightforward way to improve our simulations.

\acknowledgments

We thank Coleman Dean for helpful discussions, and the anonymous referee for constructive comments.
This research was supported by NSERC of Canada and the Faculty of Science at
the University of Alberta. The software used in this work was in part developed by 
the DOE NNSA-ASC OASCR Flash Center at the University of Chicago.
This research was enabled in part by support
provided by WestGrid (www.westgrid.ca) and Compute Canada (www.computecanada.ca). 
Computations were performed on \emph{Graham} and \emph{Cedar}. Graphics were 
developed with {\tt matplotlib} \citep{hunter2007}.



\end{document}